\documentstyle[12pt]{article}

\newcommand{\vs}[1]{\vspace{#1 mm}}

\newcommand{\beq}{\begin{equation}}
\newcommand{\eeq}{\end{equation}}
\newcommand{\beqa}{\begin{eqnarray}}
\newcommand{\eeqa}{\end{eqnarray}}
\newcommand{\nn}{\nonumber}
\newcommand{\eq}[1]{(\ref{#1})}
\newcommand{\ket}[1]{\vert\,{#1}\,\rangle}
\newcommand{\su}{{\cal U}_q (sl(2))}
\newcommand{\Lag}{{\cal L}}

\newcommand{\eps}{\epsilon}
\newcommand{\epsb}{\bar \epsilon}
\newcommand{\zb}{\bar z}

\newcommand{\ra}{\rightarrow}
\newcommand{\del}{\partial}
\newcommand{\delb}{\bar \partial}
\newcommand{\delm}{\partial_\mu}
\renewcommand{\L}{\hat L}
\newcommand{\ppp}{{\hat \phi}(z)}
\newcommand{\NP}[1]{Nucl.~ Phys.~ {\bf #1}}
\newcommand{\PL}[1]{Phys. ~Lett.~ {\bf #1}}

\newcommand{\PR}[1]{Phys.~ Rev.~ {\bf #1}}

\newcommand{\PTP}[1]{Prog.~ Theor.~ Phys.~ {\bf #1}}

\newcommand{\MPL}[1]{Mod.~ Phys.~ Lett.~ {\bf #1}}

\newcommand{\JP}[1]{J.~ Phys.~ {\bf #1}:\  Math.~Gen.~}
\begin{document}
\topmargin 0pt
\oddsidemargin 5mm
\begin{titlepage}
\setcounter{page}{0}
\begin{flushright}
revised version
\end{flushright}
\vs{13}
\begin{center}
{\large  $q$-Virasoro Operators from an Analogue of the Noether Currents
\footnote{ Extended version of the unpublished preprint
HUPD-9201.} }
\vs{20}

Haru-Tada Sato\footnote{Fellow of the Japan Society for the
Promotion of Science\\
}

\vs{5}

%

{\em Department of Physics, Hiroshima University \\
Kagamiyama 1-3-1, Higashi-Hiroshima 739, Japan}
\end{center}
\vs{10}

\begin{abstract}
We discuss the $q$-Virasoro algebra based on the arguments of the
Noether currents in a two-dimensional massless fermion theory as well as
in a three-dimensional nonrelativistic one. Some notes on
the $q$-differential operator realization and the central extension
are also included.
\end{abstract}

\vspace{1cm}

\end{titlepage}
\newpage
%
%
\section{Introduction}
\indent

Several years ago, much attention was paid to the study of $q$-deformation
of the Virasoro algebra \cite{CZ}-\cite{disc}. The $q$-deformed Virasoro
algebra was first introduced by Curtright and Zachos (CZ) as a
deformation for both commutators and structure constants \cite{CZ}. Some
other versions \cite{CILPP,DS} which can be transformed from the CZ
deformation, conformal field theoretical analogues \cite{AS}, matrix
representation \cite{NQ} and some approaches to discretized systems
\cite{disc} have also been discussed.
{}From the quantum group theoretical viewpoint \cite{DJ}, these developments of
the CZ deformation are nothing more than analogies
because a Hopf algebra structure has not been established for them.

Another type of deformed Virasoro algebra has been investigated lately
\cite{saito1}:
\beq
[L_n^{(i)},L_m^{(j)}]=\sum_{\eps=\pm 1}C^{n \hskip 8pt i}
    _{m \ \eps j}L^{(i+\eps j)}_{n+m},                       \label{EBg}
\eeq
where the structure constants are
\beq
C^{n  \ i}_{m \ j}
={[{nj-mi \over 2}]_q[(i+j)]_q \over [i]_q[j]_q} \label{EEk}
\eeq
with
\beq
[x]_q={q^x-q^{-x} \over q-q^{-1}}.
\eeq
This algebra is similar to the trigonometric algebra presented in \cite{sine},
but is different from it as is pointed out in reference \cite{CP}.
In this paper, we present some formulae based on this algebra from various
points of
view. We refer to this algebra as a (fermionic) $q$-Virasoro algebra.
It possesses a Hopf algebra structure, which is a cocommutative one.
However, its operator representation satisfies the quantum algebra $\su$
incorporating the Virasoro zero mode operator.
The Hopf algebra structure was found by
introducing an additional set of generator indices \cite{saito1} into
the differential operator representation of the CZ deformation. In this
sense, the differential operator representation plays an important role in
deriving
the situation. After this discovery, the oscillator representation \cite{CP}
and the operator product representations \cite{hsato,OS}, relevance to
a discretized Liouville model \cite{BC}, were intensively developed.
Further extensions of the $q$-Virasoro algebra have been considered: to
the supersymmetric case \cite{CP} and to more general structure constants
\cite{KS}.

In spite of these successful developements, various unsolved questions remain
about this algebra; what is the origin of this $q$-deformation or what physical
situation embodies this algebra. As for its central extension, the general
solution of the
Jacobi identity costraint equation for central extension has yet to be found.
We do
not still know a unique supersymmetric algebra which includes ghost and
superghost sectors.
Moreover, is there any further nontrivial generalization which might lead us to
a
quantum Hopf algebra structure?  These problems are important in dealing with
physical situations utilizing the $q$-Virasoro algebra.

In this paper, we would like to focus our attention on the transformation
properties of
the $q$-Virasoro algebra from a point of the classical Noether current.
In sect.2, starting from an analogy with the classical Noether current, we
define
a $q$-analogue of the canonical energy-momentum (EM) tensor.
(Throughout this paper, we often refer to an analogous object deformed by the
parameter $q$ as a $q$-object for short, e.g., $q$-EM tensor). In sect.3,
applying
the result to a two-dimensional chiral fermion theory, we define the Fourier
mode
operators of the $q$-EM tensor. It becomes clear that the mode operators
coincide with
the $q$-Virasoro generators. In this case, the $q$-EM tensor can be shown to be
related
to an analogue of conformal transformations. Next in sect. 4, from the point of
a
transformation law for a field, we discuss the relation between magnetic
translations and $q$-conformal transformations in a nonrelativistic classical
theory
under constant magnetic field. In sect.5 and 6, some brief remarks on the
differential operator associated with the $q$-conformal transformations are
added.
In the appendix, we describe a method for obtaining the central extension of
the $q$-Virasoro algebra through the Jacobi identities.

\setcounter{equation}{0}
\section{$q$-Noether current}
\indent

In this section, we discuss an analogue of the classical Noether current in
conformity with the standard Noether current in order to give a hint to
consider
the origin or the reason why the $q$-Virasoro generators take a particular form
as in \cite{BC}\cite{hsato}. Although we often refer to the commutator algebra
\eq{EBg} (including central extensions), there arises no confusion if we keep
in
mind the correspondence between the Poisson bracket and commutator.

In field theories (for any field $\phi_i$), a conserved current comes from the
invariance of the action
\beq
 \delta S=\int\delm({{\delta\Lag}\over{\delta(\delm\phi_i)}}\delta\phi_i
          -\Lag \eps^{\mu} )d^Dx,
\eeq
under an infinitesimal transformation
\beq
      x'^{\mu} = x^{\mu} - \eps^{\mu}(x),  \label{eq202}
\eeq
where $\delta\phi_i$ is the Lie derivative. If we require the invariance of
$\phi_i$ under the transformation, i.e. $\phi'_i(x')=\phi_i(x)$, the Lie
derivative
can be written in the following form
\beq
 \delta\phi_i(x)=\phi'_i(x)-\phi(x)=\eps^{\mu}\delm\phi_i(x) + O(\eps^2),
\eeq
and we get the conserved current
\beq
J^{\mu}=({\delta\Lag \over \delta(\delm\phi_i)}\del_\nu\phi_i
        -\delta^{\mu}_{\nu}\Lag )\eps^{\nu}
       =T^{\mu}_{\nu}\eps^{\nu}. \label{eq204}
\eeq
The canonical EM tensor $T^{\mu}_{\nu}$ reflects the translational
invariance of the system. In particular, the EM tensor
of the conformal field theories corresponds to the generators of
the Virasoro algebra and satisfies the conservation law
\beq
    \del_{\bar z} T(z)=\del_z {\bar T}(\zb)=0,
    \hskip 30pt T_{z\zb}=T_{\zb z}=0,       \label{eq222}
\eeq
where $T(z)=T_{zz}$ and ${\bar T}(\zb)=T_{\zb\zb}$ \cite{cft}.
Namely we obtain the generator
\beq
 L_n={1\over 2\pi i}\oint dz z^{n+1}T(z),  \label{eq224}
\eeq
which satisfies
\beq
[L_n,L_m]=(n-m)L_{n+m}+{c\over 12}(n^3-n)\delta_{n+m,0}. \label{eq223}
\eeq

When constructing a $q$-analogue of some quantity, the invariance
under the inversion $q\ra q^{-1}$ may be useful (hereafter in this
section, we denote $q$ by $Q$ ). We then insert $Q$ into the Lagrangian
so that the action should be invariant under the replacement
$Q\ra Q^{-1}$. Furthermore we assume that $Q$ is not related to the
dynamics, i.e., the action should not depend on $Q$:
\beq
 S=S(Q)=S(Q^{-1}), \hskip 30pt {d S\over d Q}=0. \label{add1}
\eeq
One of the possible ways to introduce the parameter $Q$ is in the following
change of variables in the integrand,
\beq
      S(Q)=Q^D\int\Lag (\phi_i,\del\phi_i;xQ)d^Dx,
\eeq
and
\beq
 S(Q^{-1})=Q^{-D}\int\Lag (\phi_i,\del\phi_i;xQ^{-1})d^Dx. \label{add2}
\eeq

The above requirements \eq{add1}-\eq{add2} may be reinterpreted as follows.
Hereafter we abbreviate the index $i$ of fields. The invariance of $S$ under
$Q\ra Q^{-1}$ can be understood as the
invariance under the dilatation $xQ\ra xQ^{-1}$. Namely,
this dilatation determines the $\eps^\mu(x)$ defined in \eq{eq202}
under the transformation $x\ra xQ^{-2}$
\beq
          \eps^{\mu}(x) = x^{\mu}(1-Q^{-2}).   \label{eq208}
\eeq
The invariance of $\phi$ under this transformation reads
$\phi'(x)=Q^{2\Delta}\phi(xQ^2)$, where $\Delta$ means the canonical dimension
of the field. If $\phi(x)$ is a regular function of $x$, $\phi(xQ)$ can be
expressed as an infinite series in the derivatives of $\phi(x)$
\beq
    \phi(xQ)=Q^{x\del}\phi(x).
\eeq
The Lie derivative of $\phi(x)$ is thus exactly written in the
following form
\beq
\delta\phi(x)  =Q^2 D^{(\Delta)}_{\mu}\phi(xQ)\eps^{\mu}(x),     \label{eq209}
\eeq
where $D^{\Delta}_{\mu}$ is the $Q$-derivative defined by
\beq
 (D^{(\Delta)}f)(ax)={1\over ax}{Q^{\Delta+x\del}-Q^{-\Delta-x\del} \over
Q-Q^{-1}}f(ax).
\eeq
If $Q$ is infinitesimally deviated from the unity, \eq{eq208} becomes an
infinitesimal quantity and we get the following analogue of the
conserved current,
\beq
J_Q^{\mu}=({\delta\Lag \over \delta(\delm\phi)}Q^2 D^{(\Delta)}_{\nu}\phi(xQ)
-\delta^{\mu}_{\nu}\Lag )\eps^{\nu}.            \label{eq211}
\eeq
It should be noted here that when $q=1$, \eq{eq211} coincides with the
dilatation
current
\[ j^{\mu}={\delta\Lag \over \delta(\delm\phi)}(\Delta+x^\nu\del_\nu)\phi
        -x^{\mu}\Lag \]
and can be written in the form $x^{\nu}T^{\mu}_{\nu}+\Delta AB^{\mu}C$ where
$B^\mu$
is $\del^\mu$ (or $\gamma^\mu$) for a boson (fermion) field. Thus, the
canonical
EM tensor can be obtained by putting $\Delta=0$ and dropping $x^{\nu}$ out from
$j^{\nu}$.
In the same way, we define an analogous "EM tensor" from \eq{eq211} by putting
$\Delta=0$
and dropping $\eps^{\nu}$,
\beq
 J^{\mu}_{\nu}(x)=Q^2{\delta\Lag \over \delta(\delm\phi(x))}
            D^{(0)}_{\nu}\phi(xQ)-\delta^{\mu}_{\nu}\Lag.   \label{eq216}
\eeq

This may be interpreted as a $q$-analogue of the EM tensor because it
becomes the canonical EM tensor in the limit of $Q\ra 1$.
In general, \eq{eq211} is no longer a conserved current in the case of
$Q$ being finitely deviated from unity. On the other hand in
conformal field theories, this current is trivially conserved because of the
chiral decomposition of the theories. In spite of the triviality of
conservation,
the $q$-EM tensor plays an important role in generating deformed Virasoro
algebras.
For example, in the case of a massless fermion which has the Lagrangian density
\beq
 \Lag = {1\over 2}\psi\del_{\zb}\psi
        +{1\over 2}{\bar\psi}\del_z{\bar\psi},
\eeq
the components of $J_{\mu\nu}$ are as follows;
\beqa
&\hskip 20pt J_{z\zb}=J_{\zb z}=0,                     \\
    &J(z)=J_{zz}=-{1\over 2}Q^2\psi(z)D_z\psi(zQ),      \label{eq225}\\
    &{\bar J}(\zb)=J_{\zb\zb}=-{1\over 2}Q^2{\bar\psi}
   (\zb)D_{\zb}{\bar\psi}(\zb Q).
\eeqa
The first equation of the above shows that $J_{\mu\nu}$ is traceless
$J^{\mu}_{\mu}=0$. The others indicate that $J_{zz}$ ($J_{\zb\zb}$) is
a function depending only on $z$ ($\zb$) and the conservation law of
$J_{\mu\nu}$ is
\beq
     \del_{\zb}J(z)=\del_z {\bar J}(\zb)=0.
\eeq
The $J(z)$ becomes $T(z)$ defined in \eq{eq222} in the limit $q\ra1$ and the
Fourier mode of $J(z)$ satisfies the $q$-Virasoro algebra \eq{EBg} as well as
its
classical Poisson bracket algebra. This will be clear soon in the heading of
the next section.

\setcounter{equation}{0}
\section{2-dimensional fermion current}
\indent

Now let us define the Fourier mode expansion of the $q$-EM tensor \eq{eq225}
\beq
 L_n^{(k)}={1\over 2\pi i}\oint dz z^{n+1}J^{(k)}(z)  \label{eq301}
\eeq
in which
\beq
J^{(k)}(z)={2Q^{-2} \over Q+Q^{-1}}J(zQ^{-1})
\hskip25pt\rm{with}\hskip15pt
Q=q^{k/2} \hskip 15pt (k\in{\bf Z})   \label{eq3491}
\eeq
Taking into account normal ordering in the quantum situation,
the operator \eq{eq301} satisfies the following centrally extended algebra
\cite{BC} \cite{hsato}
\beq
[L_n^{(i)},L_m^{(j)}]=\sum_{\eps=\pm 1}C^{n \hskip 8pt i}
    _{m \ \eps j}L^{(i+\eps j)}_{n+m}
    +{1\over2}C_{ij}(n)\delta_{n+m},\label{EBgg}
\eeq
with
\beq
C_{ij}(n)={1\over [i]_q[j]_q}
\sum_{k=1}^n[{(n+1-2k)i\over 2}]_q[{(n+1-2k)j\over 2}]_q\,. \label{1111}
\eeq
It is obvious that we can easily reduce the above statement to the classical
situation by omitting the central term at any time.

Next, we show that our $q$-EM tensor $J(z)$ is related to an analogy of
conformal transformations. The explicit
expression for \eq{eq301} is given in \cite{hsato}
\beq
L^{(k)}_n={-1\over2\pi i}\oint dzz^n :\psi(zq^{-k/2})
     {q^{{k\over2}z\partial}-q^{-{k\over2}z\partial}\over q^k-q^{-k}}
     \psi(z):\,\,.     \label{EBb}\eeq
This can be rewritten as
\beq
 L_n^{(k)}={-1\over 2\pi i}\oint dzz^n {q^{k(n+1)/2}\over q^k-q^{-k}}
          :\psi(z)q^{kz\del}\psi(z):.  \label{qv2}
\eeq
The factor $q^{k(n+1)/2}$ comes from the scaling of $z$.
Furthermore, $L^{(k)}$ has the following property
\beq
  L_n^{(k)}=L_n^{(-k)}\enskip,
\eeq
and thus \eq{qv2} can be symmetrized on the upper index $k$
\beq
 L_n^{(k)}={1\over 2\pi i}\oint dz {1\over2}:\psi(z)z^n
{[kz\del + k(n+1)/2]_q \over [k]_q}\psi(z):.
\label{qv12}
\eeq
Going back to the Noether current argument, the conserved charge for this
chiral
fermion theory should be
\beq
 Q_{q-Vir}={1\over 2\pi i}\oint dz :{{\delta\Lag}\over
{\delta({\bar\del}\psi)}}\delta\psi(z):, \label{qv10}
\eeq
where $\delta\psi$ means a Lie derivative for our certain particular
transformation
which might be called $q$-conformal transformations. Comparing RHSs between
\eq{qv12} and \eq{qv10}, we obtain the variation
\beq
\delta\psi(z)=z^n{[kz\del + k(n+1)/2]_q \over [k]_q}\psi(z).\label{qv5}
\eeq
This is nothing but a $q$-analogue of the
conformal transformation $\delta\psi=z^n(z\del+{1\over2}(n+1))\psi$.

\setcounter{equation}{0}
\section{3-dimensinal fermion current}
\indent

A similar situation to the previous section exists in the following
nonrelativistic
fermion field theory in two-dimensional space under constant magnetic field
\beq
\Lag_3=i\Psi^{\dagger}{\dot\Psi}-{1\over2}(D\Psi)^{\dagger}
    (D\Psi) \enskip\label{EEX}
\eeq
where $D_i=\partial_i-iA_i$. The equation of motion leads to the
Schr{\"o}dinger
equation for the Landau motion. In this system, it is known that the usual
translational invariance is modified into the so-called magnetic translational
one
\cite{mag} defined by
\beq
\Psi'(x,y)=exp(\eps b-\epsb b^\dagger)\Psi(x,y)
=T_{(\eps,\epsb)}\Psi(x,y) \enskip, \label{magtrans}
\eeq
and thus
\beq \delta \Psi = ( T_{(\eps,\epsb)} -1)\Psi(x,y) \label{mmm} \eeq
where $b$ and $b^\dagger$ are the harmonic oscillators which commute
with the Hamiltonian. In the gauge $A=(-y/2,x/2)$,
\beq
b={1\over2}{\bar w}+\del_w \enskip,\hskip 25pt
b^{\dagger}={1\over2}w-\delb_w \enskip \hskip 25pt w=x+iy
\eeq
Instead of \eq{mmm}, let us define a new transformation which is composed
of the difference between two magnetic translations
\beq
\delta\Psi={\hat {\cal L}}_n^{(k)}(w,{\bar w})\Psi
        ={{T_{(k\eps,n\epsb)}-T_{(-k\eps,n\epsb)}}
         \over{q^k-q^{-k}}}\Psi.   \label{qv11}
\eeq
The classical conserved current and charges for this transformation are given
by
\beq
J_\mu={{\delta\Lag_3}\over
      {\delta(\delm\Psi)}}\delta\Psi-\delta^{\mu}_0\Lag_3,
\eeq
\beq
{\cal L}_n^{(k)}=\int d^2w\Psi^{\dagger}(w,t)
   {T_{(k\eps,n\epsb)}-T_{(-k\eps,n\epsb)}\over q^k-q^{-k}}\Psi(w,t)\,.
\label{EEi}
\eeq
Eq.\eq{EEi} satisfies the $q$-Virasoro algebra \eq{EBg} \cite{sato}.

In order to see the similarity between \eq{qv12} and \eq{EEi},
let us compare the transformation law
\eq{qv11} with the two-dimensional relativistic case \eq{qv5}
using dimensional reduction.
We follow the method of ref.\cite{GJ} to extract holomorphic parts from
${\hat {\cal L}}_n^{(k)}(w,{\bar w})$.
After moving all ${\bar w}$ parts to the left of the $w$ parts, we
replace ${\bar w}\ra 2\del$ and $\delb\ra0$. For example,
\[
T_{(k\eps,n\epsb)}=
exp({k\eps\over2}{\bar w})exp(n\epsb\delb)exp(-{n\epsb\over2}w)
exp(k\eps\del)\hskip 10pt\ra \hskip 10pt
exp(-n\epsb w/2)exp(2k\eps\del-\eps\epsb kn/2).
\]
Further transformation is needed to see the coincidence with the
form \eq{qv5}. Taking account of the coordinate transformation from a cylinder
to the $z$-plane $ w=-{2\over\epsb}\ln z $ and of the parametrization
$ q=e^{-\eps\epsb}$,
we get a dimensionally reduced operator for ${\hat \Lag}_n^{(k)}(w,{\bar w})$
and so
\beq
\delta\Psi \ra  z^n{[kz\del_z+nk/2]_q\over[k]_q}\Psi. \label{5533}
\eeq
This expression is very similar to \eq{qv5}. This is a natural result judging
from the fact that the theory \eq{EEX} can be effectively described by a
two-dimensional massless fermion theory \cite{drop}.

In closing the section, we make a few remarks.
First, the relations among ${\hat{\cal L}}_0^{(j)}$ can be easily found in this
representation. All the ${\hat{\cal L}}_0^{(j)}$s
are written as
\beq
{\hat{\cal L}}_0^{(j)}={k^j - k^{-j} \over q^j-q^{-j}}.
\eeq
Using this relation, we obtain
\beq
{\hat{\cal L}}_0^{(2)}={1\over[2]_q}{\hat{\cal L}}_0^{(1)}
\sqrt{4+(q-q^{-1})^2{\hat{\cal L}}_0^{(1)} }
\eeq
\beq
{\hat{\cal L}}_0^{(3)}={1\over[3]_q}{\hat{\cal L}}_0^{(1)}
\{1+(q-q^{-1})[2]_q{\hat{\cal L}}_0^{(2)} \},
\eeq
and so on.
Second, one may consider the dimensional reduction
of other differentioal operator algebras for the operators $T_{(k,n)}$ or
$V_n^k$
\beq
 T_{(k,n)} \ra M_n^k \equiv z^n q^{k(z\del +n/2)} \label{qv901}
\eeq
\[
 V_n^k=2(T_{(k,n)}+T_{(-k,n)}){T_{(1,0)}-T_{(-1,0)} \over q-q^{-1}}
\]
\beq
\hskip 10pt \ra 2( M_n^k + M_n^{-k} )
          { M_0^1-M_0^{-1} \over q-q^{-1}}  \label{qv902}
\eeq
where $T_{(k,l)}$ means $T_{(k\eps,l\epsb)}$. The dimensionally reduced
differential
operators satisfy the same algebras as those before reduction
\beq
[T_{(k,n)},T_{(l,m)}]= (q^{mk-nl \over2}-q^{nl-mk \over2})T_{(k+l,n+m)}
\label{qv903}
\eeq
\beq
[V_m^j,V_n^k]= \sum_{\eps,\eta=\pm1}C^{m \hskip 8pt j}_{n \ \eps k}(\eta)
V_{m+n}^{j+\eps k+\eta}, \hskip 20pt
C^{m  \ j}_{n \ k}(r)=r[{n(j+r) -m(k+r) \over 2}]_q. \label{qv904}
\eeq
The former is called the Moyal-sine algebra \cite{sine} and the latter the
bosonic $q$-Virasoro algebra \cite{CP}.
However, this situation changes when they are put in a fermion bilinear
form. Once they are put into the fermion bilinear form like in \eq{qv2}, any
operator $O(b,b^\dagger)$ is symmetrized as $O(b,b^\dagger)-O(-b,b^\dagger)$
because of the Grassmann property of the fermion field. For example, the
insertion of the above reduced operator \eq{qv901} into the bilinear form
is equivalent to \eq{qv2} up to some normalizations, and \eq{qv2} satisfies
not the original Moyal-sine algebra \eq{qv903} but the $q$-Virasoro
algebra \eq{EBgg}. Similarly, any other algebraic relation composed of
$T_{(k,n)}$ operators becomes different from the original algebra after
inserted in the bilinear integral. Only the (fermionic) $q$-Virasoro
algebra \eq{EBgg} is preserved in this reduction procedure in the bilinear
form. In this sense, the appearance of the fermionic $q$-Virasoro algebra in
the 3-d system is nontrivial.

\setcounter{equation}{0}
\section{Differential operator algebra}
\indent

Let us consider a little more about the differential operator in \eq{5533}
(Although it is slightly different from one in \eq{qv5}, the
results after eq.\eq{eq11} do not change).
\beq
 \L_n^{(k)}=-{1\over [k]}z^n[k(z\del+{n\over 2})]\enskip.\label{eq3}
\eeq
This is known as a realization of the centerless $q$-Virasoro algebra
\eq{EBg}\cite{saito1}. It may be convenient to rewrite the operator \eq{eq3} as
\beq
  \L_n^{(k)}=-{1\over [k]} z^n \sum_{j=1}^k
        q^{(k-2j+1)(z\del+{n\over 2})}[z\del+{n\over 2}]. \label{eq4}
\eeq
Operating it on the basis
\beq
                     \ppp=z^{-h},\hskip 30pt (h\geq0)     \label{eq5}
\eeq
it  is obvious that
\beq
  \L_n^{(k)} \ppp=-z^n {[k({n\over2}-h)]\over[k]} \ppp.   \label{eq6}
\eeq
Using this relation, we obtain the following formulae
\beq
         \L_0^{(k)}\ppp={ [kh] \over [k]}\ppp   \label{qv15}
\eeq
and
\beq
\L_n^{(k)}\ppp=
{[k({n\over2}-h)]\over[k][{n\over2}-h]}\L_n^{(1)}\ppp,\label{eq8}
\eeq
\beq
[\L_n^{(i)},\L_m^{(j)}]\ppp=\sum_{\eps=\pm1}
        {[{{nj-\eps mi}\over2}][(i+\eps j)(h-{{n+m}\over2})]
        \over [i][j][h-{n+m \over2}] } \L_{n+m}^{(1)}\ppp. \label{eq9}
\eeq

In order to write the analogy with conformal primary state vectors \cite{cft},
let us introduce the following operation rule of an arbitrary polynomial $D$
of $\L_n^{(k)}$ on the 'state'
\beq
    D\ket{{\hat \phi}}\equiv\lim_{z\ra0}\ppp^{-1}D\ppp. \label{qv16}
\eeq
As a result of \eq{eq6} and \eq{qv15}, we obtain
\beqa
&\L_0^{(k)}\ket{{\hat \phi}}={ [kh] \over [k]}\ket{{\hat \phi}} \label{eq11}\\
&\L_n^{(k)}\ket{{\hat \phi}}=0 \hskip 30pt(n>0),  \label{eq12}
\eeqa
which are similar to the definition of the Virasoro primary vectors \cite{cft}.
Some other formulae for the above 'primary vectors' can be derived using
\eq{EBg},
\eq{eq8} and \eq{eq9};

\noindent (i)
\beq
\L_n^{(k)}\ket{{\hat \phi}}={[k({n\over2}-h)]
\over [k][{n\over2}-h]}\L_n^{(1)}\ket{{\hat \phi}}, \label{qv6}
\eeq
(ii) $n>0$ and $n+m\not=0$
\beq
\L_n^{(i)}\L_m^{(j)}\ket{{\hat \phi}}=\sum_{\eps=\pm1}
     {[{{nj-\eps mi}\over2}][(i+\eps j)(h-{{n+m}\over2})]
     \over [i][j][h-{n+m\over2}] } \L_{n+m}^{(1)}\ket{{\hat \phi}},
\label{eq14a}
\eeq
(iii) $n>0$, $m,l<0$, $n+m>0$ and $n+l>0$
\beq
\L_n^{(i)}\L_m^{(j)}\L_l^{(k)}\ket{{\hat \phi}}  \\
=\sum_{\eps,\eta=\pm1}
     { [{{nj-\eps mi}\over2}] [{{(n+m)k-\eta(i+\eps j)k}\over2}]
       [(i+\eps j+\eta k)(h-{{n+m+l}\over2})]
       \over [i][j][k][h-{n+m+l\over2}] }
       \L_{n+m+l}^{(1)}\ket{{\hat \phi}}\enskip .      \label{eq14b}
\eeq
Although we do not write down further formulae for higher order of $L_n^{(k)}$
\beq
\ket{{\matrix{k_1&k_2&\cdots&k_m\cr n_1&n_2&\cdots&n_m}};{\hat\phi}}
     =\prod_{j=1}^m \L_{-n_j}^{ (k_j)}\ket{\hat\phi},  \label{3976}
\eeq
it is clear that they can be also obtained streightforwardly. For example, the
eigenvalue of \eq{3976} for $L_0^{(k)}$ is given by
\beq
       {1\over[k]}[k(h+\sum_{j=1}^m n_j)].        \label{eq16}
\eeq

\setcounter{equation}{0}
\section{ Primary fields}
\indent

We rederive the formulae \eq{eq11}-\eq{eq16} presented in the previous
section from the point of a field representation. Similarly to the
Virasoro primary vectors, let us define the following primary vector
\beq
          \ket{h}=\lim_{z\ra 0}\Phi(z)\ket{0}
\eeq
in which we introduce the vacuum vector defined by
\beq
      L_n^{(k)}\ket{0}=0 \hskip 20pt (n\geq -1).
\eeq
We assume that our primary field $\Phi$ should satisfy the following
commutator
\beq
[L_n^{(k)},\Phi(z)]= {1\over [k]}z^n[k(z\del+{n\over 2}+h)]\Phi(z)
                    + A(n,h,k)z^n\Phi(z), \label{eq19}
\eeq
where $A(n,h,k)$ is an operator which satisfies the following conditions
\beqa
& \lim_{q\ra 1} A(n,h,k)=n(h-{1\over2}),         \\
           & A(n=0,h,k)=A(n,h={1\over2},k)=0.
\eeqa
It is obvious that the RHS of \eq{eq19} becomes the usual
commutator for the Virasoro primary field $z^n ( z\del+h(n+1) )\Phi(z)$ in
the limit $q\ra1$. If $h=1/2$, \eq{eq19} coincides with the case
of a massless free fermion \cite{hsato}
\beq
[L_n^{(k)},\psi(z)]={1\over [k]}z^n[kz\del+{k\over 2}(n+1)]\psi(z).
\eeq
For other value of $h$, however at present, we have not found any realization
of the same $q$-Virasoro generators even in bosonic field case.
Under these assumptions, the second term on the RHS in \eq{eq19} is not
necessary for the derivation of the formulae \eq{eq11}-\eq{eq16} as will
be shown below.

Now let us derive the formulae \eq{eq11}-\eq{eq16} from \eq{eq19}.
First, the primarity conditions \eq{eq11} and \eq{eq12} can be verified
from \eq{eq19} as
\beqa
&L_0^{(k)}\ket{h}=\lim_{z\ra0}[L_0^{(k)},\Phi(z)]
                 ={ [kh] \over [k]}\ket{h}             \\
&L_n^{(k)}\ket{h}=\lim_{z\ra0}[L_n^{(k)},\Phi(z)]
                 =0 \hskip 30pt(n>0).           \label{eq23}
\eeqa
Second, the formula \eq{qv6} is obtained as follows. Rewriting \eq{eq19}
as
\beqa
[L_n^{(k)},\Phi(z)]=&{1\over [k]}\sum_{j=1}^k
        q^{(k-2j+1)(z\del+h-{n\over 2})}
         \left( [L_n^{(1)},\Phi(z)] -A(n,h,1)z^n\Phi(z)\right) \nn \\
         & + z^nA(n,h,k)\Phi(z), \nn
\eeqa
and considering
\[
\lim_{z\ra0}[L_n^{(k)},\Phi(z)]\ket{0}={1\over [k]}\sum_{j=1}^k
   q^{(k-2j+1)(h-{n\over 2})} \lim_{z\ra0}[L_n^{(1)},\Phi(z)]\ket{0},
\]
we hence obtain
\beq
L_n^{(k)}\ket{h}
   ={[k({n\over2}-h)]\over[k][{n\over2}-h]}L_n^{(1)}\ket{h}. \label{eq25}
\eeq
Finally, the other formulae eqs.\eq{eq14a} and \eq{eq14b}
follow from \eq{EBg}, \eq{eq23} and \eq{eq25}
\beq
L_n^{(i)}L_m^{(j)}\ket{h}=\sum_{\eps=\pm1}
         {[{{nj-\eps mi}\over2}][(i+\eps j)(h-{{n+m}\over2})]
         \over [i][j][h-{n+m\over2}] } L_{n+m}^{(1)}\ket{h}
          \hskip 30pt (n>0, n+m\not=0),            \label{eq26}
\eeq
and so on.

As a simple application of the formula \eq{eq25}, we write a similar formula
for the following commutation relation \cite{hsato}
\beq
[L_n^{(j)},T^{(k)}(z)]=z^n\sum_{\eps=\pm1}{[k+\eps j]\over[j][k]}
    [{j\over2}(z\del+2)+{n\over2}(j+\eps k)]T^{(j+\eps k)}(z)
                       + C_{jk}(n)z^{n-2}            \label{eq27}
\eeq
where
\beq
  T^{(k)}(z)=\sum_{n=-\infty}^{\infty}{L_n^{(k)}\over z^{n+2}}.
\eeq
Eq.\eq{eq27} can be rewritten on the primary state using \eq{eq25}
\[
[L_n^{(j)},T^{(k)}(z)] \ket{h}
=z^n\sum_{\eps=\pm1}
{ [{j\over2}(z\del+2)+{n\over2}(j+\eps k)] \over [j] }
{ [(k+\eps j)(h+1+{1\over2}z\del)] \over [k(h+1+{1\over2}z\del)] }
T^{(k)}(z)\ket{h}
\]
\beq
+ C_{jk}(n)z^{n-2}\ket{h}.   \label{qv100}
\eeq
Integrating this formula, we can verify the algebra \eq{EBgg} on the state
\[
\hskip -80pt [L_n^{(i)},L_m^{(j)}]\ket{h}
={1\over2\pi i} \oint dz z^{m+1}[L_n^{(i)},T^{(j)}(z)]\ket{h}
\]
\[
\hskip 50pt=\sum_{\eps=\pm 1}
           {[{\eps nj-mi \over 2}][i+\eps j] \over [i][\eps j]}
           L^{(i+\eps j)}_{n+m}\ket{h}+C_{ij}(n)\ket{h}.
\]
All the formula in this section are explicitly verified in the case
of massless fermion with $h=1/2$.

\setcounter{equation}{0}
\section{ Conclusions}
\indent

The $q$-Virasoro algebra may be understood as a natural extention
of the quantum algebra $\su$ \cite{add1} in spite of possessing an ordinary
Hopf algebra structure and thus it is different from the $W_{1+\infty}$ algebra
\cite{add2} in this point. We have discussed various features of it in fermion
systems and in the differential operator realization starting from deforming
the
canonical EM tensor.

We have shown that the $q$-EM tensor defined in sect.2 generates the
$q$-Virasoro
algebra in two-dimensional chiral fermion theory.
However, we would like to mention here that there remain a few points which
should be investigated as further problems.
First, the $q$-EM tensor for higher dimensions is not conserved in the sense
either of ordinary divergence equation or of $q$-derivative equation.
Second, we have not found any relevance of the $q$-EM tensor to the bosoinic
$q$-Virasoro algebra \eq{qv904} \cite{BC,KS,CP} in contrast with the success
for the
fermionic case
\beqa
           &J(z)=Q^2\del\phi(z)D_z\phi(zQ),       \\
&{\bar J}(\zb)=Q^2\del_{\zb}{\bar\phi}(\zb)D_{\zb}{\bar\phi}(\zb Q).
\eeqa
This situation might be related to the fact that an isomorophism between the
fermionic
$q$-Virasoro algebra and the bosonic one has not been found.
{}From this point of view, improvement of the $q$-EM tensor of this paper must
be an
interesting topic as a further work.

We have attempted to formulate the $q$-Virasoro algebra as an expression of
invariance under a deformed transformation, which is called the $q$-conformal
transformation. We have considered the deformation of a conserved current in an
undeformed field theory. In distinction to this case, there may exist a more
fundamental
approach for analyzing a conserved current after constructing a deformed field
theory
as well. Finally, we expect that our analysis may point the way to a more
suitable
formulation of the $q$-conformal transformation law of a fermion/boson field
and to a clarification of the meaning of the $q$-Virasoro algebra.

\vspace{1cm}
\noindent
{\em Acknoledgements}

This work was done while visiting the theory group of M{\"u}nchen University.
The author would like to thank Prof. J. Wess for kind hospitality,
Prof. R. Sasaki and Dr. N. Aizawa for stimulating communications at
an early stage of this work. Thanks are also due to Prof. A. Solomon
for reading the manuscript.

\newpage
\setcounter{equation}{0}
\appendix
\section{Central extension}
\indent

Although the central extension was found in \cite{CP} taking account of
normal ordering of fermion oscillators, we present another method to determine
the
central extension through solving the Jacobi identity conditions.
The central extension should be found from the Jacobi identity;
\beq
\sum_{\eps=\pm 1}C^{m  \hskip 6pt j}_{\hskip 3pt l \hskip 7pt \eps k}
C(n,m+l;i,j+\eps k) + \rm{cycl. perms.} =0,      \label{A12}
\eeq
where both indices $(i,j,k)$ and $(n,m,l)$ should be permuted
simultaneously. $C(n,m;i,j)$ is defined by
\beq
[L_n^{(i)},L_m^{(j)}]=\sum_{\eps=\pm 1}C^{n \hskip 8pt i}_{m \ \eps j}
       L^{(i+\eps j)}_{n+m}+C(n,m;i,j),     \label{A13}
\eeq
and it satisfies the following relations
\beqa
 &C(n,m;i,j)=C(n,m;|i|,|j|),     \label{A14a} \\
 &C(n,m;i,j)=-C(m,n;j,i).        \label{A14}
\eeqa
Although eq.\eq{A12} has the solution
\[
  C(n,m;i,j)=C^{n  \ i}_{m \ j} + C^{n  \hskip 10pt i}_{m \ -j},
\]
it is trivially eliminated by the translation
$ L_n^{(i)} \hskip 10pt \ra\hskip 10pt L_n^{(i)}-1$.
As it is difficult to find the general solution of eq.\eq{A12}, we
impose the condition
\beq
     C(n,0;i,j) = C(1,-1;i,j) = 0.       \label{A15}
\eeq
The above conditions reflect those satisfied by the ordinary Virasoro
algebra. In order to find a solution of \eq{A12}, we fix some redundant
variables to be $n=-l-1$, $m=1$, $k=i$ and $j=ai$ ($a$ is an integer).
Using \eq{A14a} and \eq{A14}, eq.\eq{A12} becomes
\[
\sum_\eps \{
C^{1  \ ai}_{\hskip 3pt l \ \eps i}C(-l-1,l+1;i,(a+\eps)i)
\]
\beq
\hskip 40pt + C^{-l-1 \hskip 6pt i}_{\hskip 10pt 1 \hskip 8pt a\eps i}
  C(l,-l;i,(a+\eps)i) \}=0.            \label{A16}
\eeq
Assuming
\beq
      C(n,m;i,j)=C(n,m;j,i),      \label{A17}
\eeq
we can immediately find the simplest solution putting $a=1$ in
\eq{A16},
\beq
 C^{1  \ i}_{l \hskip 6pt i}C(l+1,-l-1;2i,i)=
 C^{-l-1 \ i}_{\hskip 10pt 1  \hskip 12pt i} C(l,-l;i,2i). \label{A18}
\eeq
The solution of this equation is
\beq
C(n,m;i,2i)={[{i\over 2}(n+1)][{i\over 2}n][{i\over 2}(n-1)]
  \over [3{i\over 2}][i][{i\over 2}] }{\hat C}(i)\delta_{n+m,0}.
\label{A19}
\eeq
The ${\hat C}(i)$ is $C(2,-2;i,2i)$ and corresponds to the central
charge $c/2$ in the limit $q\ra 1$.
We can adjust the normalization as in \eq{1111} choosing
\beq
{\hat C}(i)={[i/2]\over[2i]}2c
\eeq
or making use of null state condition for the conformal weight
$1/2$ state \cite{hsato}.

All other solutions should be recursively determined from
\eq{A16} as follows. First fix the values of
$C(n,-n;j,j)$ and $C(n,-n;2j,j)$ as the initial conditions of the
recursive equations. Then all of the $C(n,-n;i+2j,j)$ can be
determined after $C(n,-n;i,j)$ $(1\leq i\leq 2j)$ are obtained.
Hence, all the solutions of \eq{A16} will be found recursively.


%
\end{document}